\documentclass[final,5p,times,twocolumn,nopreprintline]{elsarticle}
\usepackage{amsmath,amssymb,amsfonts,dsfont}
\usepackage{graphicx}
\usepackage{slashed}
\usepackage{booktabs,tabulary}

\newcommand{\beq}{\begin{equation}}
\newcommand{\eeq}{\end{equation}}
\newcommand{\beqa}{\begin{eqnarray}}
\newcommand{\eeqa}{\end{eqnarray}}
\newcommand{\Lagr}{\mathcal{L}}
\newcommand{\gA}{g_\text{A}}
\newcommand{\Fpi}{F_\pi}
\newcommand{\mpi}{M_{\pi}}

\newcommand{\meta}{M_{\eta}}

\newcommand{\diff}{\text{d}}
\newcommand{\eps}{\epsilon}
\newcommand{\M}{\mathcal{M}}
\newcommand{\MNR}{\mathcal{M}_{\rm NR}}

\newcommand{\Order}{\mathcal{O}}
\newcommand{\Op}{\mathcal{O}}
\newcommand{\qq}{\mathbf{q}}

\newcommand{\PP}{\mathbf{P}}
\newcommand{\KK}{\mathbf{K}}
\newcommand{\spin}{\mathbf{S}}
\newcommand{\vv}{\mathbf{v}^\bot}
\newcommand{\xx}{\mathbf{x}}
\newcommand{\sig}{\boldsymbol{\sigma}}
\newcommand{\ttau}{\boldsymbol{\tau}}

\newcommand{\GeV}{\,\text{GeV}}
\newcommand{\MeV}{\,\text{MeV}}

\newcommand{\muu}{m_u}
\newcommand{\md}{m_d}
\newcommand{\mpp}{m_p}
\newcommand{\mn}{m_n}
\newcommand{\mN}{m_N}
\newcommand{\mA}{m_A}
\newcommand{\mc}{m_\chi}

\newcommand{\nf}{N_\text{f}}
\renewcommand{\arraystretch}{1.2}

\allowdisplaybreaks[1]

\begin{document}

\renewcommand{\theequation}{\arabic{equation}}

\begin{frontmatter}

\title{Chiral power counting of one- and two-body currents 
in direct detection of dark matter}

\author[IKP,EMMI]{Martin Hoferichter}
\author[IKP,EMMI]{Philipp Klos}
\author[IKP,EMMI]{Achim Schwenk}

\address[IKP]{Institut f\"ur Kernphysik, 
Technische Universit\"at Darmstadt, 64289 Darmstadt, Germany}
\address[EMMI]{ExtreMe Matter Institute EMMI, 
GSI Helmholtzzentrum f\"ur Schwerionenforschung GmbH, 64291 Darmstadt, Germany}

\begin{abstract}
We present a common chiral power-counting scheme for vector,
axial-vector, scalar, and pseudoscalar WIMP--nucleon interactions, and
derive all one- and two-body currents up to third order in the chiral
expansion.  Matching our amplitudes to non-relativistic effective
field theory, we find that chiral symmetry predicts a hierarchy
amongst the non-relativistic operators. 
Moreover, we identify interaction channels where two-body currents that previously have not been accounted for 
become relevant. 
\end{abstract}


\end{frontmatter}

\section{Introduction}

Elucidating the nature of dark matter is one of the most pressing
challenges in contemporary particle physics and astrophysics.  Still,
one of the dominant paradigms rests on a weakly-interacting massive
particle (WIMP), such as the neutralino in supersymmetric extensions
of the standard model (SM). A WIMP can be searched for at colliders,
in annihilation signals, or in direct-detection experiments, where the
recoil energy deposited when the WIMP scatters off nuclei is
measured. Recent years have witnessed an impressive increase in
sensitivity, e.g., from XENON100~\cite{Aprile:2012nq},
LUX~\cite{Akerib:2013tjd}, and SuperCMDS~\cite{Agnese:2014aze}, which
will further improve dramatically with the advent of ton-scale
detectors, XENON1T~\cite{Aprile:2012zx} and
LZ~\cite{Malling:2011va}. In the absence of a signal, direct-detection
experiments provide more and more stringent constraints on the
parameter space of WIMP candidates. To derive these constraints and to
interpret a future signal, it is mandatory that the nucleon matrix
elements and the nuclear structure factors, which are required when
transitioning from the SM to the nucleon to the nucleus level,
be calculated systematically and incorporate what we know about QCD.

Effects at the level of the nucleus can be described by an effective
field theory (EFT) whose degrees of freedom are non-relativistic (NR)
nucleon and WIMP fields~\cite{Fan:2010gt,Fitzpatrick:2012ix}. This
NREFT has been recently used in an analysis of direct-detection
experiments~\cite{Schneck:2015eqa}. In this approach, scales related
to the spontaneous breaking of chiral symmetry of QCD are integrated
out, with the corresponding effects subsumed into the coefficients of
the EFT. In the context of nuclear forces, such an EFT is called
pionless EFT. To derive limits on the WIMP parameter space,
information from QCD has then to be included in the analysis in a
second step.

Alternatively, one can start directly from chiral EFT (ChEFT) to
incorporate the QCD constraints from chiral
symmetry~\cite{Prezeau:2003sv,Cirigliano:2012pq,Menendez:2012tm,Klos:2013rwa,Baudis:2013bba,Cirigliano:2013zta,KlosMaster,Vietze:2014vsa},
which makes predictions for the hierarchy among one- and two-body
currents. Based on ChEFT, scalar and axial-vector two-body currents
were recently considered in~\cite{Cirigliano:2012pq}
and~\cite{Menendez:2012tm,Klos:2013rwa}, respectively. Moreover,
lattice QCD can be used to constrain the couplings of two-body
currents~\cite{Beane:2013kca}.

The goal of this Letter is to combine vector, axial-vector, scalar, and
pseudoscalar interactions in a common chiral power counting, collect
all relevant one- and two-body matrix elements, and match the result
onto NREFT. This combines our knowledge of QCD at low energies: the
one-body matrix elements correspond to the standard decomposition into
form factors, while the two-body
scalar~\cite{Prezeau:2003sv,Cirigliano:2012pq},
vector~\cite{Pastore:2008ui,Kolling:2009iq,Kolling:2011mt}, and
axial-vector~\cite{Park:2002yp,KlosMaster} currents
have been calculated as well, the vector current even at one-loop
order. Here, we combine these results for their application in direct
detection, extending the axial-vector two-body currents to
finite momentum transfer and generalizing to the three-flavor case
where appropriate. By matching to the NREFT, we find that the chiral
symmetry of QCD predicts a hierarchy among the different operators and
that two-body currents can be as important as one-body currents in
some channels.

\section{Effective Lagrangian and kinematics}
\label{sec:EFT}

We start from the following dimension-$6$ and -$7$ effective
Lagrangian for the interaction of the WIMP $\chi$, assumed to be a SM
singlet, with the SM fields~\cite{Goodman:2010ku}
\begin{align}
\label{Lagr}
 \Lagr_{\chi}&=\frac{1}{\Lambda^3}\sum_q\Big[C_{q}^{SS}\bar \chi \chi \,m_q\bar q q
 +C_{q}^{PS}\bar \chi i\gamma_5 \chi \,m_q\bar q q \notag\\
 &\qquad+C_{q}^{SP}\bar \chi \chi \,m_q\bar q i\gamma_5 q
 +C_{q}^{PP}\bar \chi i\gamma_5 \chi \,m_q\bar q i\gamma_5 q\Big]\notag\\
 &+\frac{1}{\Lambda^2}\sum_q\Big[C_q^{VV}\bar\chi\gamma^\mu\chi \,\bar q\gamma_\mu q
 +C_q^{AV}\bar\chi\gamma^\mu\gamma_5\chi\, \bar q\gamma_\mu q\notag\\
 &\qquad+C_q^{VA}\bar\chi\gamma^\mu\chi\, \bar q\gamma_\mu\gamma_5 q
 +C_q^{AA}\bar\chi\gamma^\mu\gamma_5\chi\, \bar q\gamma_\mu\gamma_5 q\Big]\notag\\
 &+\frac{1}{\Lambda^2}\sum_q\Big[C_q^{TT}\bar\chi\sigma^{\mu\nu}\chi\, \bar q\sigma_{\mu\nu} q
 +\tilde C_q^{TT}\bar\chi\sigma^{\mu\nu}i\gamma_5\chi\, \bar q\sigma_{\mu\nu} q\Big]\notag\\
 &+\frac{1}{\Lambda^3}\Big[C^S_{g}\bar \chi\chi\, \alpha_sG_{\mu\nu}^aG^{\mu\nu}_a
 +C^P_{g}\bar \chi i\gamma_5\chi\, \alpha_sG_{\mu\nu}^aG^{\mu\nu}_a\notag\\
 &\qquad+\tilde C^S_{g}\bar \chi\chi\, \alpha_sG_{\mu\nu}^a\tilde G^{\mu\nu}_a
 +\tilde C^P_{g}\bar \chi i\gamma_5\chi\, \alpha_sG_{\mu\nu}^a\tilde G^{\mu\nu}_a\Big],
\end{align}
where the Wilson coefficients $C_i$ parameterize the effect of new
physics associated with the scale $\Lambda$ (organizing the interactions
in this way assumes $\Lambda$  
to be much larger than the typical QCD scale of $1\GeV$).
To render the scalar and
pseudoscalar matrix elements renormalization-scale invariant we
included explicitly the quark masses $m_q$ in the definition of the
respective operators.  We further assumed $\chi$ to be a Dirac fermion
(in the Majorana case, $C_q^{VV}=C_q^{VA}=C_q^{TT}=0$), and defined
the dual field strength tensor as
\beq
\tilde G^{\mu\nu}_a=\frac{1}{2}\eps^{\mu\nu\lambda\sigma}G^a_{\lambda\sigma},
\eeq
with sign convention $\eps^{0123}=+1$.
Compared to the operator basis used in~\cite{Hill:2014yxa} we do not
include the dimension-$8$ operators related to the traceless part of
the QCD energy-momentum tensor. 
As shown in~\cite{Hill:2014yxa}, these operators become relevant for heavy WIMPs
and contribute to spin-independent interactions, decreasing significantly the single-nucleon
contribution.
Finally, we will ignore the tensor
operators in~\eqref{Lagr} and concentrate on the chiral predictions
for the $V,A,S,P$ channels.

The kinematics for the WIMP--nucleon scattering process are taken as
\beq
N(p)+\chi(k)\to N(p')+\chi(k'),
\eeq
the momentum transfer is defined as
\beq
\label{momenta}
q=k'-k=p-p',\qquad q^2= t,
\eeq
and the pion, $\eta$, nucleon, nucleus, and WIMP masses will be denoted by $\mpi$, $\meta$, $\mN$,
$\mA$, and $\mc$, respectively (Dirac spinors are normalized to~$1$).
We will also need
\beq
P=p+p',\qquad K=k+k'.
\eeq

The cross section differential with respect to momentum transfer for
the elastic WIMP--nucleus scattering process in the laboratory frame can be
expressed as
\beq
\label{cross_sec}
\frac{\diff \sigma}{\diff \qq^2}=\frac{1}{8\pi v^2(2J+1)}\sum_\text{spins}|\MNR|^2+\Order\big(\qq^0\big),
\eeq
with nucleus spin $J$, WIMP velocity $v$, and NR amplitude $\MNR$ defined as
\beq
\M=2\mA 2\mc\MNR+\Order\big(\qq^2\big),
\eeq
where $\M$ is the relativistic scattering amplitude. 
In the Majorana case, \eqref{cross_sec} receives an additional
factor of $4$.

\section{Chiral power counting}
\label{sec:counting}

\begin{table*}
\centering
\renewcommand{\arraystretch}{1.3}
\begin{tabular}{clcccccc}
\toprule
& Nucleon & & $V$ & & & $A$ &\\
WIMP & & $t$ & & $\xx$ & $t$ & & $\xx$ \\\midrule
& $1$b & $0$ & & $1+2$ & $2$ & & $0+2$\\
$V$ & $2$b & $4$ & & $2+2$ & $2$ & & $4+2$\\
& $2$b NLO & -- & & -- & $5$ & & $3+2$\\\midrule
& $1$b & $0+2$ & & $1$ & $2+2$ & & $0$\\
$A$ & $2$b & $4+2$ & & $2$ & $2+2$ & & $4$\\
& $2$b NLO & -- & & -- & $5+2$ & & $3$\\
\bottomrule
\end{tabular}
\qquad\qquad
\begin{tabular}{clcc}
\toprule
& Nucleon & $S$ & $P$ \\
WIMP & & & \\\midrule
& $1$b & $2$ & $1$\\
$S$ & $2$b & $3$ & $5$\\
& $2$b NLO & --& $4$ \\\midrule
& $1$b & $2+2$ & $1+2$\\
$P$ & $2$b & $3+2$ & $5+2$\\
& $2$b NLO & -- & $4+2$\\
\bottomrule
\end{tabular}
\renewcommand{\arraystretch}{1.0}
\caption{Left: leading chiral order of time ($t$) and space ($\xx$) 
components of the WIMP and nucleon currents for vector and
axial-vector interactions, for one-body ($1$b) and two-body ($2$b)
operators. For the axial-vector nucleon operator, terms involving
vertices from the NLO chiral Lagrangian (indicated by ``$2$b NLO'') need
to be included (see main text for details). The second number (``$+2$'') 
refers to the additional suppression originating from the NR expansion
of the WIMP spinors, if momentum over WIMP mass is counted in the same
way as for the nucleon mass. Right: leading chiral order of the WIMP and
nucleon currents for scalar and pseudoscalar interactions.}
\label{tab:VASP}
\end{table*}

We use the standard chiral power counting~\cite{Gasser:1983yg,Gasser:1984gg}
\beq
\partial=\Order\big(p\big),\qquad m_q=\Order\big(p^2\big),\qquad
a_\mu,v_\mu=\Order(p),
\eeq
with axial-vector and vector sources $a_\mu$ and $v_\mu$. 
The velocity distribution in dark matter halo models indeed suggests to count the momentum transfer $q \lesssim \mpi$ 
as $\Order(p)$~\cite{Cirigliano:2012pq}.
In the
baryon sector we depart from the standard counting in chiral
perturbation theory (ChPT) and adopt the more conventional ChEFT
assumption (see, e.g.,~\cite{Weinberg:1991um,Ordonez:1995rz,Epelbaum:2008ga}) for the
scaling of relativistic corrections
\beq
\frac{\partial}{\mN}=\Order\big(p^2\big).
\eeq
This counting is appropriate for a break-down scale around $500\MeV$.
As far as the WIMP is concerned, a chiral counting is only required for
the NR expansion of the spinors. We assume the same counting as in the
nucleon case, but display the corresponding additional powers
explicitly. If $\mc \gtrsim\mN$, the suppression will be more
pronounced, for $\mpi\lesssim\mc \lesssim\mN$ the counting should be
adapted, and for even smaller $\mc$ the naive counting breaks down.

For most of the channels it suffices to consider the leading-order
Lagrangian to determine at which chiral order a given contribution
starts. For the one-body matrix elements higher orders are subsumed
into the nucleon form factors, which are obtained by their chiral
expansion or could be taken from phenomenology. In this work, we
consider all contributions up to $\Order(p^3)$. Since the leading
two-body terms start at $\Order(p^2)$, this leaves the possibility
that the next-to-leading-order (NLO) pion--nucleon Lagrangian involving
the low-energy constants $c_i$~\cite{Bernard:1992qa} could be
required, and this is indeed the case for the spatial component of the
axial-vector current~\cite{Menendez:2012tm,Klos:2013rwa} (indicated by ``$2$b NLO'' in
Table~\ref{tab:VASP}). In the same channel, $NN$ contact terms
$d_i$~\cite{Cohen:1995cc} enter. We define both $c_i$ and $d_i$ in the
conventions of~\cite{Park:2002yp} (with dimensionless $c_6$ and $c_7$).

As a preview of our results, the leading chiral orders of one- and
two-body currents for time and space components of the axial-vector
and vector currents, as well as for the scalar and pseudoscalar
operators, are listed in Table~\ref{tab:VASP}. The suppression by two
powers (``$+2$'') originating from the WIMP spinors is displayed
separately. In the following sections, we give results for all one-
and two-body currents involved in Table~\ref{tab:VASP}.

\section{Nuclear matrix elements}
\label{sec:nuc}

\subsection{Scalar}

At zero momentum transfer the scalar couplings of the heavy quarks
$Q=c,b,t$ can be determined from the trace anomaly of the QCD
energy-momentum tensor~\cite{Shifman:1978zn}
\begin{align}
\label{EM_trace}
\theta^\mu_{\ \mu}&=\sum_q m_q\bar q q+\frac{\beta_\text{QCD}}{2g_s}G^a_{\mu\nu}G_a^{\mu\nu},\qquad
\langle N|\theta^\mu_{\ \mu}|N\rangle=\mN,
\notag\\
\frac{\beta_\text{QCD}}{2g_s}&=-\bigg(11-\frac{2\nf}{3}\bigg)\frac{\alpha_s}{8\pi}+\Order\big(\alpha_s^2\big).
\end{align}
For $\nf=3$ active flavors, one obtains
\beq
\langle N|m_Q\bar Q Q|N\rangle=-\frac{\alpha_s}{12\pi}\langle N\big|G_{\mu\nu}^aG^{\mu\nu}_a\big|N\rangle=\mN f_Q^N,
\eeq
where 
\beq
\label{scalar_couplings}
f_Q^N=\frac{2}{27}\bigg(1-\sum_{q=u,d,s}f_q^N\bigg),\qquad \mN f_q^N=\langle N|m_q\bar q q|N\rangle.
\eeq
Therefore, at leading order in $\alpha_s$ the effect of integrating out the heavy quarks can be absorbed into a redefinition of $C^S_{g}$
\beq
C'^S_{g}=C^S_{g}-\frac{1}{12\pi}\sum_{Q=c,b,t}C^{SS}_{Q}.
\eeq
For the $u$- and $d$-quarks the couplings are intimately related to the pion--nucleon $\sigma$-term $\sigma_{\pi N}$~\cite{Crivellin:2013ipa}
\begin{align}
\label{fq_res}
f_u^N&=\frac{\sigma_{\pi N}(1-\xi)}{2\mN}+\Delta f_u^N,\quad
&f_d^N&=\frac{\sigma_{\pi N}(1+\xi)}{2\mN}+\Delta f_d^N,
\end{align}
with $\xi=\frac{\md-\muu}{\md+\muu}=0.36\pm
0.04$~\cite{Colangelo:2010et} and corrections $\Delta f_{u,d}^N$
related to the strong proton--neutron mass difference via the low-energy
constant $c_5$. For the strange quark, the most accurate determination
comes from lattice QCD~\cite{Junnarkar:2013ac}.
The above $\Order(\alpha_s)$ analysis may not be accurate enough for the charm quark, 
see~\cite{Kryjevski:2003mh,Vecchi:2013iza,Hill:2014yxa} for a study of higher orders in $\alpha_s$. 

This analysis generalizes to finite $t$ if one defines
\begin{align}
\mN f_q^N(t)&=\langle N(p')|m_q\bar q q|N(p)\rangle, \qquad \theta_0^N(t)=\langle N(p')|\theta^\mu_{\ \mu}|N(p)\rangle, \notag\\
f_Q^N(t)&=\frac{2}{27}\bigg(\frac{\theta_0^N(t)}{\mN}-\sum_{q=u,d,s}f_q^N(t)\bigg),
\end{align}
and replaces $f_q^N\to f_q^N(t)$, $f_Q^N\to f_Q^N(t)$ accordingly.

The chiral expansion of $\sigma_{\pi N}$ starts with
\beq
\sigma_{\pi N}=-4c_1\mpi^2+\Order\big(p^3\big),
\eeq
in line with the $\Order(p^2)$ listed in Table~\ref{tab:VASP} for
the scalar one-body current. Note, however, that the power $2$ does not imply a
momentum-dependent coupling in this case, but a quark-mass suppression. As far as the
$t$-dependence is concerned, the slope of the scalar form factors is
dominated by $\pi\pi$ scattering, which is known to not be
adequately described by ChPT, but to require a reconstruction based on
dispersion relations~\cite{Gasser:1990ap,Hoferichter:2012wf,Ditsche:2012fv}.
The $t$-dependence generated by other sources but light-quark scalar form factors 
was shown to be higher order in the chiral expansion in~\cite{Cirigliano:2012pq}.

Defining
\beq
f_N(t)=\frac{m_N}{\Lambda^3}\bigg(\sum_{q=u,d,s}C^{SS}_{q}f_q^N(t)-12\pi f^N_Q(t)C'^S_{g}\bigg),
\eeq
the NR one-body matrix element for the scalar channel becomes\footnote{The nucleon spinors include isospin indices according to $\chi_{s'}^\dagger f_N(t)\chi_s\equiv\frac{1}{2}\chi_{s'}^\dagger\Big[ \big(f_p(t)+f_n(t)\big)\mathds{1}+\big(f_p(t)-f_n(t)\big)\tau^3\Big]\chi_s$. The Wilson coefficients match onto the conventions of~\cite{Engel:1992bf}
by means of the identification $f_N(0)=\sqrt{2}G_F c_0$.}
\beq
\M_{1,\text{NR}}^{SS}= \chi_{r'}^\dagger\chi_r \chi_{s'}^\dagger f_N(t)\chi_s,
\eeq
where $\chi_{r,s}$ ($\chi_{r',s'}$) are NR spinors for the incoming (outgoing) WIMP and nucleon, respectively.
$\M_{1,\text{NR}}^{PS}$ is of higher chiral order since the NR reduction of $\gamma_5$ produces a term $-\sig\cdot\qq/(2\mc)$, which we count as $\Order(p^2)$ for $\mc \gtrsim \mN$.

\subsection{Vector}

The decomposition of the vector current at the quark level reads
\beq
\langle N(p')|\bar q \gamma^\mu q|N(p)\rangle = \langle N'|\gamma^\mu F_1^{q,N}(t)-\frac{i\sigma^{\mu\nu}}{2\mN}q_\nu F_2^{q,N}(t)|N\rangle,
\eeq
where the sign of the Pauli term is due to the convention in~\eqref{momenta}. To obtain a flavor decomposition of the vector current, one 
usually assumes isospin symmetry (corrections can again be calculated in ChPT~\cite{Kubis:2006cy}):
\beq
F_i^{u,p}(t)=F_i^{d,n}(t),\qquad F_i^{d,p}(t)=F_i^{u,n}(t),\qquad F_i^{s,p}(t)=F_i^{s,n}(t).
\eeq
In this way, one obtains
\begin{align}
F_i^{u,p}(t)&=F_i^{d,n}(t)=2F_i^{\text{EM},p}(t)+F_i^{\text{EM},n}(t)+F_i^{s,N}(t),\notag\\
F_i^{d,p}(t)&=F_i^{u,n}(t)=F_i^{\text{EM},p}(t)+2F_i^{\text{EM},n}(t)+F_i^{s,N}(t),
\end{align}
with electromagnetic form factors $F_i^{\text{EM},N}(t)$. 
At vanishing momentum transfer this defines the vector couplings
\beq
\label{fV}
\langle N|\bar q \gamma^\mu q|N\rangle=f_{V_q}^N\langle N|\gamma^\mu|N\rangle,\qquad
f_{V_u}^p=f_{V_d}^n=2f_{V_d}^p=2f_{V_u}^n=2.
\eeq
Corrections to~\eqref{fV} can be worked out in terms of magnetic moments $\mu_N=Q_N+\kappa_N$, electric radii $\langle r_E^2\rangle^N$, as well as strangeness moments $\mu_N^s=\kappa_N^s$ and radii $\langle r_{E,s}^2\rangle^N$, explicitly
\begin{align}
F_1^{u,p}(t)&=2+2\bigg(\frac{\langle r_E^2\rangle^p}{6}-\frac{\kappa_p}{4\mpp^2}\bigg)t
+\bigg(\frac{\langle r_E^2\rangle^n}{6}-\frac{\kappa_n}{4\mn^2}\bigg)t\notag\\
&+\bigg(\frac{\langle r_{E,s}^2\rangle^N}{6}-\frac{\kappa_s}{4\mN^2}\bigg)t+\Order\big(t^2\big),\notag\\
F_1^{d,p}(t)&=1+\bigg(\frac{\langle r_E^2\rangle^p}{6}-\frac{\kappa_p}{4\mpp^2}\bigg)t
+2\bigg(\frac{\langle r_E^2\rangle^n}{6}-\frac{\kappa_n}{4\mn^2}\bigg)t\notag\\
&+\bigg(\frac{\langle r_{E,s}^2\rangle^N}{6}-\frac{\kappa_s}{4\mN^2}\bigg)t+\Order\big(t^2\big),\notag\\
F_1^{s,N}(t)&=\bigg(\frac{\langle r_{E,s}^2\rangle^N}{6}-\frac{\kappa_N^s}{4\mN^2}\bigg)t+\Order\big(t^2\big), 
\notag\\
F_2^{u,N}&=\kappa_N+\Order(t), \qquad F_2^{d,N}=-\kappa_N-\kappa_N^s+\Order(t), \notag\\
F_2^{s,N}&=\kappa_N^s+\Order(t),
\end{align}
with the Sachs form factors
\begin{align}
G_E^N(t)&=F_1^N(t)+\frac{t}{4\mN^2} F_2^N(t)=Q_N+\frac{\langle r_E^2\rangle^N}{6}t+\Order\big(t^2\big),\notag\\
G_M^N(t)&=F_1^N(t)+F_2^N(t)=\mu_N\Bigg(1+\frac{\langle r_M^2\rangle^N}{6}t\Bigg)+\Order\big(t^2\big).
\end{align}

The NR one-body matrix elements involving a nucleon vector current are
\begin{align}
\M_{1,\text{NR}}^{VV}&=\chi_{r'}^\dagger\chi_r\chi_{s'}^\dagger\bigg[f_1^{V,N}(t)-\frac{\qq}{4\mN^2}\cdot\Big(\qq-i\sig\times\PP\Big)f_2^{V,N}(t)\bigg]\chi_s\notag\\
&+\frac{1}{2\mc}\chi_{r'}^\dagger\Big[\KK+i\sig\times\qq\Big]\chi_r\cdot
\frac{1}{2\mN} \chi_{s'}^\dagger i\sig\times\qq f_2^{V,N}(t)\chi_s,\notag\\
\M_{1,\text{NR}}^{AV}&=\frac{1}{2\mc}\chi_{r'}^\dagger\sig\cdot\KK \chi_r\chi_{s'}^\dagger f_1^{A,N}(t)\chi_s\notag\\
&\hspace{-10pt}-\chi_{r'}^\dagger\sig\chi_r\cdot\frac{1}{2\mN}\chi_{s'}^\dagger\bigg[\Big(\PP-i\sig\times\qq\Big)f_1^{A,N}(t)
-i\sig\times\qq f_2^{A,N}(t)\bigg]\chi_s,
\end{align}
where
\begin{align}
f_i^{V,N}(t)=\frac{1}{\Lambda^2}\sum_{q=u,d,s}C^{VV}_{q}F_i^{q,N}(t),\notag\\
f_i^{A,N}(t)=\frac{1}{\Lambda^2}\sum_{q=u,d,s}C^{AV}_{q}F_i^{q,N}(t).
\end{align}

\subsection{Axial vector}

The decomposition of the axial-vector current at the quark level reads (see, e.g.,~\cite{Bernard:2001rs,Schindler:2006it})
\begin{align}
\langle N(p')|\bar q \gamma^\mu\gamma_5 q|N(p)\rangle &= \langle N'|\gamma^\mu\gamma_5 G_A^{q,N}(t)-\gamma_5\frac{q^\mu}{2\mN} G_P^{q,N}(t)\notag\\
&-\frac{i\sigma^{\mu\nu}}{2\mN}q_\nu\gamma_5 G_T^{q,N}(t)|N\rangle.
\end{align}
$G_T^{q,N}(t)$ corresponds to a second-class current~\cite{Weinberg:1958ut}, i.e., it violates $G$-parity, and will be ignored in the following. At vanishing momentum transfer only $G_A^{q,N}$ contributes. Its coefficients are conventionally defined as
\beq
\langle N(p)|\bar q \gamma^\mu\gamma_5 q|N(p)\rangle=\Delta q^N \langle N|\gamma_\mu\gamma_5|N\rangle,
\eeq
and isospin symmetry is assumed
\beq
\label{DeltaqN}
\Delta u^p=\Delta d^n,\qquad \Delta u^n=\Delta d^p,\qquad \Delta s^p=\Delta s^n.
\eeq
The combinations
\begin{align}
a_3^p&=-a_3^n=\Delta u^p-\Delta d^p=\gA,\notag\\
a_8^N&=\Delta u^N+\Delta d^N-2\Delta s^N=3F-D,
\end{align}
are determined by the axial charge of the nucleon in the case of $a_3$, or can be inferred from semileptonic hyperon decays for $a_8$, yielding $D\approx 0.8$, $F\approx 0.46$. 
The third combination
\beq
\Delta\Sigma^N=\Delta u^N+\Delta d^N+\Delta s^N
\eeq
is related to the spin structure function of the nucleon, it is not a scale-independent quantity. At $Q^2=5\GeV^2$ and $\Order(\alpha_s^2)$
the following values were obtained in~\cite{Airapetian:2006vy}
\begin{align}
\Delta u^p&=0.842\pm 0.012,\qquad \Delta d^p = -0.427\pm 0.013, \notag\\
\Delta s^p&= -0.085\pm0.018.
\end{align}
Besides the coefficients at zero also the momentum dependence of the flavor combinations
\begin{align}
A_\mu^3&=\bar Q \gamma_\mu\gamma_5\frac{\lambda^3}{2}Q=\frac{1}{2}\big(\bar u \gamma_\mu\gamma_5 u-\bar d\gamma_\mu\gamma_5 d\big),\notag\\
A_\mu^8&=\bar Q \gamma_\mu\gamma_5\frac{\lambda^8}{2}Q=\frac{1}{2\sqrt{3}}\big(\bar u \gamma_\mu\gamma_5 u+\bar d\gamma_\mu\gamma_5 d-2\bar s \gamma_\mu\gamma_5 s\big),
\end{align}
can be analyzed in SU($\nf$) ChPT, but due to the anomalously broken U$(1)_A$ current this is not the case for the isoscalar component. 
One obtains 
\begin{align}
 \langle N(p')|A_\mu^3|N(p)\rangle &= \langle N'|\bigg(\gamma^\mu\gamma_5 G_A^3(t)-\gamma_5\frac{q^\mu}{2\mN} G_P^3(t)\bigg)\frac{\tau^3}{2}|N\rangle,\notag\\
 \langle N(p')|A_\mu^8|N(p)\rangle &= \langle N'|\bigg(\gamma^\mu\gamma_5 G_A^8(t)-\gamma_5\frac{q^\mu}{2\mN} G_P^8(t)\bigg)\frac{1}{2}|N\rangle,
\end{align}
with leading-order results
\begin{align}
\label{GA_LO}
G_A^3(t)&=g_A,\qquad G_A^8(t)=\frac{3F-D}{\sqrt{3}}\equiv g_A^8,\notag\\
G_P^3(t)&=-\frac{4\mN^2 g_A}{t-\mpi^2},\qquad G_P^8(t)=-\frac{4\mN^2 g_A^8}{t-\meta^2}.
\end{align}
Empirically, the momentum dependence of $G_A^3(t)$, extracted from neutrino scattering off nucleons and charged-pion electroproduction, follows a dipole fit
\beq
G_A^3(t)=\frac{\gA}{(1-t/M_A^2)^2},
\eeq
with mass parameter $M_A$ around $1\GeV$~\cite{Bernard:2001rs,Schindler:2006it}. Since for general $t$ the flavor structure cannot be inverted without additional input for the singlet component, we decompose the quark sum according to\footnote{At vanishing momentum transfer this equation maps onto the notation of~\cite{Engel:1992bf} by means of $\sum_qC_q^{AA}\Delta q^N=\sqrt{2}G_F\Lambda^2\frac{1}{2}\big(a_0+a_1\tau^3\big)$.}
\beq
\sum_q C_q^{AA}G_{A,P}^{q,N}(t)=C_0^{AA} G_{A,P}^0(t)+C_3^{AA} G_{A,P}^3(t)\tau^3+C_8^{AA} G_{A,P}^8(t),
\eeq
with
\begin{align}
C_0^{AA}&=\frac{1}{3}\Big[C_u^{AA}+C_d^{AA}+C_s^{AA}\Big],\qquad
C_3^{AA}=\frac{1}{2}\Big[C_u^{AA}-C_d^{AA}\Big],\notag\\
C_8^{AA}&=\frac{\sqrt{3}}{6}\Big[C_u^{AA}+C_d^{AA}-2C_s^{AA}\Big],
\end{align}
and define
\beq
g_{A,P}^N(t)=\frac{1}{\Lambda^2}\Big[C_0^{AA} G_{A,P}^0(t)+C_3^{AA} G_{A,P}^3(t)\tau^3+C_8^{AA} G_{A,P}^8(t)\Big].
\eeq
In terms of these quantities, the NR amplitude reads
\beq
\M_{1,\text{NR}}^{AA}=-\chi^\dagger_{r'}\sig\chi_r \cdot
\chi_{s'}^\dagger\bigg[\sig g_A^N(t)-\frac{\qq}{4\mN^2}\sig\cdot\qq g_P^N(t) \bigg]\chi_s.
\eeq
Similarly, for the $VA$ channel we define
\beq
h_{A,P}^N(t)=\frac{1}{\Lambda^2}\Big[C_0^{VA} G_{A,P}^0(t)+C_3^{VA} G_{A,P}^3(t)\tau^3+C_8^{VA} G_{A,P}^8(t)\Big],
\eeq
to obtain
\begin{align}
\M_{1,\text{NR}}^{VA}&=\chi^\dagger_{r'}\chi_r
\frac{1}{2\mN}\chi_{s'}^\dagger\bigg[\sig\cdot\PP h_A^N(t)-\frac{\qq\cdot\PP}{4\mN^2}\sig\cdot\qq h_P^N(t) \bigg]\chi_s\\
&\hspace{-11pt}-\frac{1}{2\mc}\chi^\dagger_{r'}\Big[\KK+i\sig\times\qq\Big]\chi_r\cdot
\chi_{s'}^\dagger
\bigg[\sig h_A^N(t)-\frac{\qq}{4\mN^2}\sig\cdot\qq h_P^N(t) \bigg]\chi_s.\notag
\end{align}

\subsection{Pseudoscalar}

The pseudoscalar matrix element is usually parameterized as
\beq
\langle N(p')|m_q\bar q i\gamma_5 q|N(p)\rangle = \langle N'|\mN G_5^{q,N}(t) i\gamma_5|N\rangle.
\eeq
By means of the Ward identity
\beq
\label{axial_Ward}
\sum_q\partial_\mu \bar q\gamma^\mu\gamma_5 q=\sum_q 2i m_q\bar q\gamma_5 q-\frac{\alpha_s\nf}{4\pi}G_{\mu\nu}^a\tilde G^{\mu\nu}_a,
\eeq
the corresponding form factor $G_5^{q,N}(t)$ follows from $G_A^{q,N}(t)$ and $G_P^{q,N}(t)$, except for the singlet component, where
the anomaly does not drop out,
\beq
G_5^{i}(t)=G_A^{i}(t)+\frac{t}{4\mN^2}G_P^{i}(t),\qquad i=3,8.
\eeq
Accordingly, we have
\begin{align}
 \M_{1,\text{NR}}^{SP}&=\chi^\dagger_{r'}\chi_r
 \frac{i}{2}\chi_{s'}^\dagger \sig\cdot\qq g_{5}^N(t)\chi_s,\notag\\
 \M_{1,\text{NR}}^{PP}&=\frac{1}{2\mc}\chi_{r'}^\dagger\sig\cdot\qq\chi_r
 \frac{1}{2}\chi_{s'}^\dagger \sig\cdot\qq h_{5}^N(t)\chi_s,
\end{align}
where
\begin{align}
 g_5^N(t)&=\frac{1}{\Lambda^2}\Big[C_3^{SP} G_5^3(t)\tau^3+C_8^{SP} G_{5}^8(t)\Big],\notag\\
 h_5^N(t)&=\frac{1}{\Lambda^2}\Big[C_3^{PP} G_5^3(t)\tau^3+C_8^{PP} G_{5}^8(t)\Big].
\end{align}

\section{Two-body currents}
\label{sec:MEC}

\subsection{Scalar}

The scalar meson-exchange currents, involving both pion and $\eta$ contributions, have been considered before in~\cite{Prezeau:2003sv,Cirigliano:2012pq}. The full expression reads
\begin{align}
 \M_{2,\text{NR}}^{SS}&=-\chi_{r'}^\dagger\chi_r\bigg(\frac{\gA}{2\Fpi}\bigg)^2f_\pi\mpi^2
 \chi_{s_1'}^\dagger\chi_{s_2'}^\dagger\ttau_1\cdot\ttau_2
X_{12}^\pi
\chi_{s_1}\chi_{s_2}\notag\\
&-\chi_{r'}^\dagger\chi_r\bigg(\frac{\gA}{2\Fpi}\bigg)^2\bigg(\frac{4\alpha-1}{\sqrt{3}}\bigg)^2f_\eta\meta^2
\chi_{s_1'}^\dagger\chi_{s_2'}^\dagger
X_{12}^\eta
\chi_{s_1}\chi_{s_2},
\end{align}
where
\beq
X_{12}^i=\frac{\sig_1\cdot\qq_1\,\sig_2\cdot\qq_2}{\big(\qq_1^2+M_i^2\big)\big(\qq_2^2+M_i^2\big)},\qquad i=\pi,\eta,
\eeq
pion decay constant $\Fpi=92.2\MeV$~\cite{PDG}, 
$\chi_{s_i}$ ($\chi_{s_i'}$) denote NR spinors for the incoming (outgoing) nucleons, with momenta $p_i$ ($p_i'$), $q_i=p_i'-p_i$,
$\alpha=F/(D+F)$, and the Wilson coefficients are collected in
\beq
f_\pi=\frac{1}{\Lambda^3}\sum_{q=u,d} C^{SS}_{q}f_q^\pi,\qquad
f_\eta=\frac{1}{\Lambda^3}\sum_{q=u,d,s} C^{SS}_{q}f_q^\eta,
\eeq
with scalar meson couplings
\beq
f_u^\pi=\frac{\muu}{\muu+\md}=0.32\pm0.03,\qquad f_d^\pi=\frac{\md}{\muu+\md}=0.68\pm0.03,
\eeq
and
\begin{align}
f_u^\eta&=\frac{1}{3}\frac{\muu}{\muu+\md}\frac{\mpi^2}{\meta^2}=(6.9\pm 0.4)\times 10^{-3},\notag\\
f_d^\eta&=\frac{1}{3}\frac{\md}{\muu+\md}\frac{\mpi^2}{\meta^2}=(14.7\pm 0.4)\times 10^{-3},\notag\\
f_s^\eta&=\frac{2}{3}\frac{M_{K^0}^2+M_{K^+}^2-\mpi^2}{\meta^2}=1.05.
\end{align}
One particular feature of the scalar two-body currents is that they cannot be written as a correction to the one-body coupling $f_N$, 
since the scalar couplings of pions and $\eta$ mesons probe a different combination of Wilson coefficients~\cite{Cirigliano:2012pq}. 
For this reason, even in the isospin limit they cannot be parameterized in terms of a single coupling $c_0$ as conventionally done for the one-body currents, see e.g.~\cite{Engel:1992bf}.

\subsection{Vector}

The only two-body vector current up to $\Order(p^3)$ appears in the $AV$ channel
\begin{align}
\label{2b:AV}
 \M_{2,\text{NR}}^{AV}&=-\frac{2}{\Lambda^2}C_3^{AV}\bigg(\frac{\gA}{2\Fpi}\bigg)^2\chi_{r'}^\dagger\sig\chi_r\cdot \chi_{s_1'}^\dagger\chi_{s_2'}^\dagger i\big[\ttau_1\times\ttau_2\big]^3
 \bigg[\frac{\sig_1\cdot \qq_1\,\sig_2}{\qq_1^2+\mpi^2}\notag\\
 &-\frac{\sig_2\cdot \qq_2\,\sig_1}{\qq_2^2+\mpi^2}+\big(\qq_1-\qq_2\big)X_{12}^\pi\bigg]
\chi_{s_1}\chi_{s_2}.
\end{align}
While the nucleon vector current itself has been studied in detail before~\cite{Pastore:2008ui,Kolling:2009iq,Kolling:2011mt},
the present application to direct detection is new.

In fact, there are neither terms with $i=8$ nor $\eta$ contributions
to $i=3$. The reason for this can be traced back to the operator
structure of the chiral Lagrangian: the coupling to the vector current
occurs via a commutator $[v_\mu,\phi]$ of vector source and meson
matrix. Expanded in Gell-Mann matrices, this leaves SU(3) structure
factors $f^{3ij}$ and $f^{8ij}$, and the only non-trivial ones, apart
from the direct couplings to the nucleon that led to~\eqref{GA_LO},
reduce to the SU(2) subset $\epsilon^{ijk}$.

\subsection{Axial vector}

The axial-vector two-body currents are
\begin{align}
\M_{2,\text{NR}}^{AA}&=\frac{1}{\Lambda^2}C_3^{AA}\chi^\dagger_{r'}\sig\chi_r\cdot
\chi_{s_1'}^\dagger\chi_{s_2'}^\dagger 
\Bigg\{\Bigg[\frac{\gA}{\Fpi^2}\big[\ttau_1\times\ttau_2\big]^3\bigg[\frac{c_6}{4}\sig_1\times\qq\notag\\
&\qquad+c_4\bigg(1-\frac{\qq}{\qq^2+\mpi^2}\qq\cdot\bigg)\sig_1\times\qq_2\bigg]\frac{\sig_2\cdot \qq_2}{\qq_2^2+\mpi^2}\notag\\
&+\frac{2\gA}{\Fpi^2}\tau_2^3\bigg[2c_1\mpi^2\frac{\qq}{\qq^2+\mpi^2}\notag\\
&\qquad+c_3\bigg(\qq_2-\frac{\qq}{\qq^2+\mpi^2}\qq\cdot \qq_2\bigg)\bigg]\frac{\sig_2\cdot \qq_2}{\qq_2^2+\mpi^2}\notag\\
&+2d_1\tau_1^3\bigg(\sig_1-\frac{\sig_1\cdot \qq\,\qq}{\qq^2+\mpi^2}\bigg)\Bigg]+\big(1 \leftrightarrow 2\big)\notag\\
&+2d_2\big[\ttau_1\times\ttau_2\big]^3\big(\sig_1\times\sig_2\big)\bigg(1-\cdot\qq\frac{\qq}{\qq^2+\mpi^2}\bigg)\Bigg\}
\chi_{s_1}\chi_{s_2},
\end{align}
where the terms that do not contain an explicit $\qq$-dependence ($\qq=-\qq_1-\qq_2$) and the $c_6$-term are taken from~\cite{Park:2002yp},
while the finite-$\qq$ pion-pole corrections were derived in~\cite{KlosMaster}. The $AA$ two-body current as in~\cite{Park:2002yp} has been 
applied in the calculation of structure factors for spin-dependent scattering in~\cite{Menendez:2012tm,Klos:2013rwa}, 
whereas the two-body current in the $VA$ channel, 
\begin{align}
\label{2b:VA}
\M_{2,\text{NR}}^{VA}&=-\frac{1}{\Lambda^2}C_3^{VA}\frac{\gA}{2\Fpi^2}\chi^\dagger_{r'}\chi_r
\chi_{s_1'}^\dagger\chi_{s_2'}^\dagger 
\Bigg\{i\big[\ttau_1\times\ttau_2\big]^3\frac{\sig_2\cdot \qq_2}{\qq_2^2+\mpi^2}\notag\\
&\qquad+\big(1 \leftrightarrow 2\big)\Bigg\}
\chi_{s_1}\chi_{s_2}, 
\end{align}
has not been considered before.

For similar reasons as in the vector case there are no $i=8$ or $\eta$
contributions from the leading-order Lagrangian. In principle, one
could calculate corrections from the NLO SU(3) Lagrangian, in analogy
to the SU(2) result for $\M_{2,\text{NR}}^{AA}$. However, there is a
large number of poorly-known low-energy constants
(see~\cite{Frink:2004ic} or~\cite{Mai:2009ce} for the matching to
SU(2)), which would severely limit the predictive power.

Finally, due to the derivative in the Ward identity~\eqref{axial_Ward}, there are no pseudoscalar two-body currents at $\Order(p^3)$.

\section{Matching to NREFT}
\label{sec:match}

Next, we express our results in terms of the operator basis from~\cite{Fitzpatrick:2012ix}
\begin{align}
\label{Op_Wick}
 \Op_1&=\mathds{1}, & \Op_2&=\big(\vv\big)^2, & \Op_3&=i\spin_N\cdot (\qq\times\vv),\notag\\ 
  \Op_4&=\spin_\chi\cdot \spin_N, & \Op_5&=i\spin_\chi\cdot \big(\qq\times\vv\big), & \Op_6&=\spin_\chi\cdot \qq\,\spin_N\cdot \qq,\notag\\
  \Op_7&=\spin_N\cdot \vv, & \Op_8&=\spin_\chi\cdot \vv, & \Op_9&=i\spin_\chi\cdot \big(\spin_N\times \qq\big),\notag\\
  \Op_{10}&=i\spin_N\cdot \qq, & \Op_{11}&=i\spin_\chi\cdot \qq, & &
\end{align}
where $\spin=\sig/2$ and the velocity is defined as
\beq
\label{def_vel}
\vv=\frac{\KK}{2\mc}-\frac{\PP}{2\mN}.
\eeq
We find the relations
\begin{align}
\label{matching}
 \M_{1,\text{NR}}^{SS}&= \chi_{r'}^\dagger\chi_{s'}^\dagger \Op_1 f_N(t)\chi_r\chi_s,\notag\\
 \M_{1,\text{NR}}^{SP}&=\chi^\dagger_{r'}\chi_{s'}^\dagger \Op_{10} g_{5}^N(t)\chi_r\chi_s,\notag\\ 
 \M_{1,\text{NR}}^{PP}&=\frac{1}{\mc}\chi_{r'}^\dagger \chi_{s'}^\dagger \Op_6h_{5}^N(t)\chi_r\chi_s,\notag\\
 \M_{1,\text{NR}}^{VV}&=\chi_{r'}^\dagger\chi_{s'}^\dagger\bigg[\Op_1\Big(f_1^{V,N}(t)+\frac{t}{4\mN^2}f_2^{V,N}(t)\Big)
 +\frac{1}{\mN}\Op_3 f_2^{V,N}(t)\notag\\
 &\qquad+\frac{1}{\mN\mc}\Big(t \Op_4+\Op_6\Big)f_2^{V,N}(t)\bigg]\chi_r\chi_s,\notag\\
\M_{1,\text{NR}}^{AV}&=\chi^\dagger_{r'}\chi_{s'}^\dagger\bigg[2\Op_8 f_1^{A,N}(t)
+\frac{2}{\mN}\Op_9\Big(f_1^{A,N}(t)+f_2^{A,N}(t)\Big)\bigg]\chi_r\chi_s,\notag\\
\M_{1,\text{NR}}^{AA}&=\chi^\dagger_{r'}\chi^\dagger_{s'}\bigg[
-4\Op_4 g_A^N(t)+\frac{1}{\mN^2}\Op_6 g_P^N(t)\bigg]\chi_r\chi_s,\notag\\
\M_{1,\text{NR}}^{VA}&=\chi^\dagger_{r'}\chi^\dagger_{s'}\bigg[-2\Op_7+\frac{2}{\mc}\Op_9\bigg]h_A^N(t)\chi_r\chi_s.
\end{align}
This shows that as a result of QCD effects, the operators in the NREFT
are not independent. For example, both axial and pseudoscalar operators 
combine in the nuclear matrix element $\M_{1,\text{NR}}^{AA}$.
In addition, up to $\Order(p^3)$ only $8$ of the
$11$ operators of~\eqref{Op_Wick} are present. However, because
$\M_{1,\text{NR}}^{PS}$ itself enters only at $\Order(p^4)$, they are
mapped onto $7$ amplitudes, so that the relations cannot be inverted.
This is because $\M_{1,\text{NR}}^{AV}$ and $\M_{1,\text{NR}}^{VA}$
involve the three operators $\Op_{7-9}$. This implies that some operators, e.g.\ $\Op_6$,
can be isolated by having a particular quark-level interaction, but this is not possible in general,
as demonstrated by the example of $\Op_{7-9}$.
If we retain subleading
corrections in the NR expansion of the spinors, the missing operators
appear, accompanied by additional combinations: $\Op_{11}$ in terms of
$\M_{1,\text{NR}}^{PS}$, $\Op_2$ and $\Op_5$ in
$\M_{1,\text{NR}}^{VV}$, $\Op_3\Op_8$ in $\M_{1,\text{NR}}^{AV}$, and
$\Op_7\Op_8$ in $\M_{1,\text{NR}}^{AA}$.

In the limit where $\mc$ becomes (significantly) larger than the
nucleon mass also $\M_{1,\text{NR}}^{PP}$ should be dropped, as well
as the $1/\mc$ suppressed terms in $\M_{1,\text{NR}}^{VV}$ and
$\M_{1,\text{NR}}^{VA}$. In contrast, all two-body currents up to
$\Order(p^3)$ are independent of $\mc$. They appear in the $SS$, $AV$, $AA$, and $VA$ channels. 

We stress that the above discussion merely pertains to the mapping of operator structures, it does not
take into account the evolution of the scale dependence that is required when matching the coefficients 
of a pionless theory, valid for scales below the pion mass, and ChEFT, defined at chiral scales. 
This involves also effects related to the limitations of the ``Weinberg'' counting scheme applied here~\cite{Valderrama:2014vra}, and would have 
to be taken into account in the matching relations required for translating NREFT coefficients to the QCD scale.
In addition, there may be effects from operator mixing, originating from the interplay between the nucleon-spin dependence in the ChEFT WIMP--nucleon scattering operator and that in the high-momentum part of the ChEFT $NN$ potential, which would also have to be considered when evolving NREFT operators to the QCD scale.

\section{Summary and discussion}
\label{sec:summary}

In this Letter, we have developed the constraints that chiral symmetry of QCD imposes
on the nuclear matrix elements that can enter in dark matter direct detection.
We provide explicit expressions for one- and two-body
currents in WIMP--nucleus scattering for vector, axial-vector, scalar,
and pseudoscalar interactions up to third order in the chiral
expansion. The chiral power counting, summarized in
Table~\ref{tab:VASP}, shows that at this order there are two-body
currents that have not been considered and may be of similar or
greater importance than some of the one-body operators, see~\eqref{2b:AV} and~\eqref{2b:VA}. Moreover, the matching to NREFT
shows that not all allowed one-body operators appear at this chiral
order and that the operators in the NREFT are not independent.

The chiral power counting applies to the one- and two-nucleon level.
In nuclei, the different interactions can lead to a coherent response
that scales with the number of nucleons in the nucleus or to a
single-particle-like response. In a next step, we will evaluate the
nuclear structure factors, including the contributions from two-body
currents, and provide a set of response functions for the analysis of
direct-detection experiments. This will also allow us to assess how 
constructive or destructive the interference of operators based on 
the constraints provided by chiral symmetry proves to be.

\section*{Acknowledgements}

We thank J.\ Men\'endez for useful discussions, and B.\ Kubis, U.-G.~Mei{\ss}ner, and M.\ J.\ Savage for comments on the manuscript. This work was
supported by BMBF ARCHES, the DFG through Grant No.\ SFB 634, and the
ERC Grant No.\ 307986 STRONGINT.

\end{document}